\documentclass[a4paper,12pt]{extarticle}
\usepackage{graphicx}
\usepackage{rotating}
\usepackage{amssymb}
\usepackage{mathptmx}
\usepackage[numbers]{natbib}
\usepackage[english]{babel}

\bibpunct{}{}{,}{s}{}{,}

\begin{document}

\newcommand{\hdblarrow}{H\makebox[0.9ex][l]{$\downdownarrows$}-}

\begin{center}
   \textbf{\LARGE{Fast Diffusion Process in Quenched hcp Dilute Solid $^3$He-$^4$He Mixture}}
\\[0.5cm]
\textsl{Ye.O. Vekhov, A.P. Birchenko, N.P. Mikhin, and E.Ya. Rudavskii}
\\[0.5cm]
\textsl{\small{B.Verkin Institute for Low Temperature Physics and Engineering of the National Academy of Sciences of Ukraine, 47 Lenin ave., Kharkov 61, 61103, Ukraine\\ 
{vekhov@ilt.kharkov.ua}
}}



\end{center}

\begin{abstract}

The study of phase structure of dilute $^3$He - $^4$He solid mixture of different quality is performed by spin echo NMR technique. The diffusion coefficient is determined for each coexistent phase. Two diffusion processes are observed in rapidly quenched (non-equilibrium) hcp samples: the first process has a diffusion coefficient corresponding to hcp phase, the second one has huge diffusion coefficient corresponding to liquid phase. That is evidence of liquid-like inclusions formation during fast crystal growing. It is established that these inclusions disappear in equilibrium crystals after careful annealing.



\end{abstract}

\section{Introduction}

Last time the interest has sharply increased for searching the conditions for realization supersolidity phenomenon in solid $^4$He \cite{Balibar.2008,Prokofev.2007}, when the crystalline order combines with superfluidity. In spite of the great number of experimental and theoretical investigations in this area, the consensus has not been attained yet. For the present, it has been determined well that observing effects strongly depend on the growing conditions and annealing degree of helium crystals. The special modeling which was conducted from the first principles by Monte-Carlo method, showed that in the perfect hcp $^4$He crystal the supersolidity effects cannot appear \cite{Prokofev.2007, Boninsegni.2006}. 

The most authors connect such effects in solid $^4$He at low temperatures with the disorder in helium samples. Possible kinds of the disorder may be the defects, grain boundaries \cite{Sasaki.2006}, glass phase, or liquid inclusions \cite{Boninsegni.2006,Balatsky.2007,Grigorev.2007}. Also, the possible interpretation \cite{Balibar2.2008} of the experiments on flow the superfluid helium through the solid helium \cite{Ray.2008} show the essential role of the liquid channels, which may exist in the solid helium up to the ultralow temperatures. In this connection, the experiments which allow to identify the kind of the disorder, for example, in rapidly grown helium crystals, interesting. These data can be obtained by nuclear magnetic resonance (NMR). Whereas for its realization the nuclei of $^3$He are necessary, we deal hereafter with the samples of not pure $^4$He  but with dilute $^3$He-$^4$He mixture.

Since NMR technique allows to measure diffusion coefficient in different coexisting phases and difference of diffusion coefficients in liquid and solid helium are several orders of the magnitude then such an experiment may answer the question -- whether liquid inclusions are formed in solid helium under very rapid crystal growing. The aim of present work is to elucidate this problem. We detect, by NMR technique, the presence of liquid phase in solid helium samples grown in different conditions and also establish the influence of annealing effect on character of diffusion processes. 

\section{Experimental procedure}
\label{Method}

The crystals were grown by the capillary blocking method from initial helium gas mixture with a 1\% of $^3$He concentration. The copper cell of cylindrical form with inner diameter of 8~mm and length of 18~mm has the NMR coil glued to the inner surface of the cell. The pressure and temperature variations of the sample in the cell were controlled by two capacitive pressure gauges fixed to the both cylinder ends and by two resistance thermometers attached to the cold finger of the cell with sensitivities about 1~mbar and 1~mK, respectively.

Two series of crystals under the pressure above 33~bar were studied. The first one ("low quality crystals") was prepared by quick step-wise cooling from the melting curve down to the lowest temperature (1.27~K) without any special thermal treatment. To improve the crystal quality of the second series ("high quality crystals") a special three-stage thermal treatment was used: annealing at the melting curve, thermocycling in single phase regions and annealing in the HCP single phase region near the melting curve \cite{Birchenko.2006}. The criterions of crystal quality are, first, constancy of the pressure with time under constant temperature which is closed to melting and, second, reaching the pressure minimum under thermal cycling.

\begin{figure}[ht]
                \centerline{\includegraphics[scale=1.0]{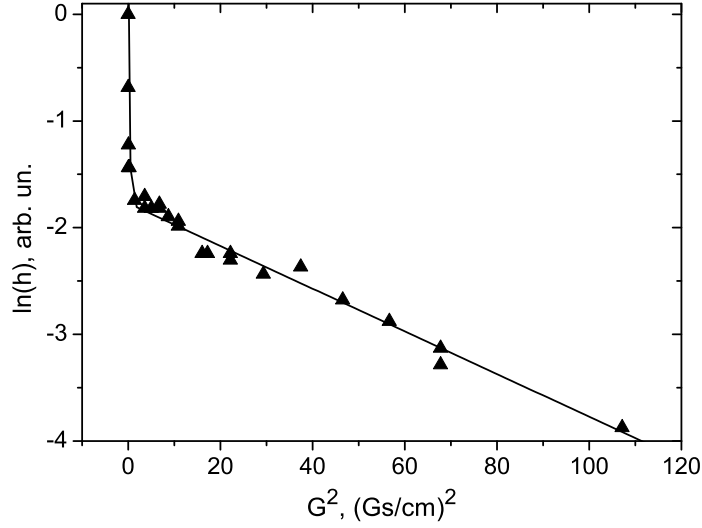}}
                \caption{\small{Diffusion decay of spin-echo in  hcp crystal on the melting curve, 2.10~K, 45.0~bar. Points -- experimental SE data for $\tau_1=80$~ms, $\tau_2=1$~s, $D_l=1 \cdot 10^{-5}$~cm$^2$/s, $D_h=1.8 \cdot 10^{-8}$~cm$^2$/s; the solid line -- fitting to Eq.~(\ref{1}).}}
                 \label{fig_mc}
\end{figure}

The spin diffusion coefficient was determined with the help of the pulsed NMR technique at a frequency of $7.12$~MHz. The Carr-Purcell ($CP$) spin-echo method \cite{Carr.1954,Hann.1950} was used with a 90$^\circ$-$\tau$-180$^\circ$ sequence of probe pulses as well as the method of stimulated echo ($SE$) with the sequence of three probes pulses 90$^\circ$-$\tau_1$-90$^\circ$-$\tau_2$-90$^\circ$ were applied to the nuclear system of the sample. 

Generally, if a few phases do coexist in the sample, the echo amplitude $h$  for $CP$ is given by 

\begin{equation}
\frac{h}{h_0}=\sum_{i}\alpha_i\, \exp\left(-\frac{2}{3}\gamma^2G^2\tau ^3D_i\right)
\label{1}
\end{equation}
 and for $SE$
\begin{equation}
\frac{h}{h_0}=\sum_{i}\alpha_i\, \exp\left[-\gamma^2G^2D_i\tau_1^2
\left(\tau_2-\frac{\tau_1}{3}\right)\right]
\label{2}
\end{equation}
 where $h_0$ is the maximal amplitude of a  echo amplitude at $G\to 0$, $G$ is the magnetic field gradient, $\gamma$ is a gyromagnetic ratio,  index $i$ numerates   coexisting phases with the diffusion coefficients $D_i$, $\alpha_i$ is the relative content of the $i$-th phase in the sample. 

One can choose duration parameters  $\tau$, $\tau_1$, and $\tau_2$ in order to get the strongest $h(G^2)$ dependence and to single out $D_i$ fitting parameter.
It should be emphasized that spin-diffusion coefficient $D$ measurement 
  was just the method to identify a thermodynamical phases by their typical $D$ value. Neither contribution of $^3$He atoms in a phase transition processes nor even the dynamics of different phase's ratio could be tracking because of too long spin-lattice relaxation times.

\section{Results}
The typical results of NMR measurements for diffusion coefficients in two-phase sample on the melting curve are presented in Fig.~\ref{fig_mc} in $h(G^2)$ scale. There are two slopes for the data obtained which correspond to two different diffusion coefficients. Experimental data analysis according to Eq.~(\ref{1}) gives for curve piece with sharp slope $D_l=1 \cdot 10^{-5}$~cm$^2$/s which corresponds to diffusion in liquid phase \cite{Garwin.1959} and for curve piece with mildly slope $D_h=1.8 \cdot 10^{-8}$~cm$^2$/s which corresponds to diffusion in hcp phase \cite{Allen.1982,Mikheev.1983}. The phase ratio is $\alpha_l/\alpha_h\sim 3/4$.

\begin{figure}[ht]
                \centerline{\includegraphics[scale=1.0]{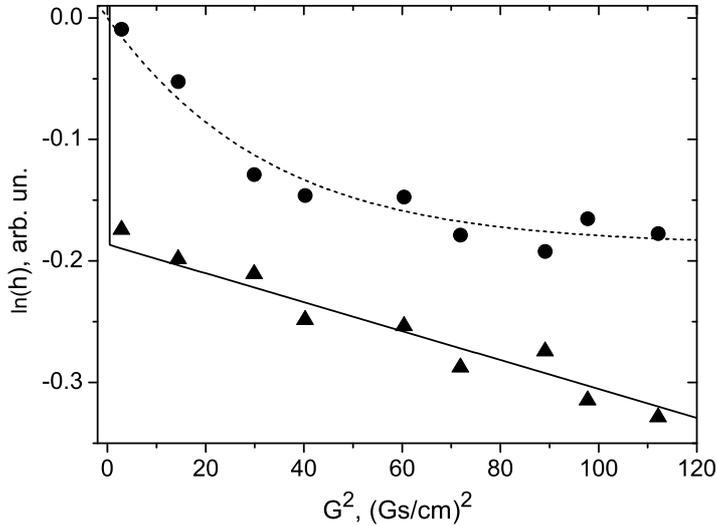}}
                \caption{\small{Diffusion decay of spin-echo in quenched hcp crystal, 1.3~K, 35.0~bar. Solid circles -- $CP$ data for $\tau=25$~ms, $D_l=1 \cdot 10^{-4}$~cm$^2$/s; solid triangles -- $SE$ data for $\tau_1=14$~ms, $\tau_2=20$~s, $D_h=8 \cdot 10^{-10}$~cm$^2$/s; the dashed line -- fitting to Eq.~(\ref{1}); the solid line -- fitting to Eq.~(\ref{2}).}}
                 \label{fig_quenched}
\end{figure}

Then this sample was rapidly cooled down to 1.3~K in the hcp region. The results of NMR measurements are shown in Fig.~\ref{fig_quenched}. The presence of significant contribution ($\sim 17\%$) of phase with fast diffusion coefficient ($D_l=1 \cdot 10^{-4}$~cm$^2$/s) was unexpected. This fact can be interpreted as existence of liquid-like inclusions in hcp matrix which were apparently quenched from the melting curve. Such a situation was visually observed in pure $^4$He in Refs.~[1,4,15,16].
The liquid droplets formation was also observed by NMR technique in 1\% $^3$He-$^4$He mixture under bcc and hcp phases coexistence \cite{Mikhin.2001,Polev.2002}. Note that this effect was observed in all three low-quality samples studied.

\begin{figure}[ht]
                \centerline{\includegraphics[scale=1.0]{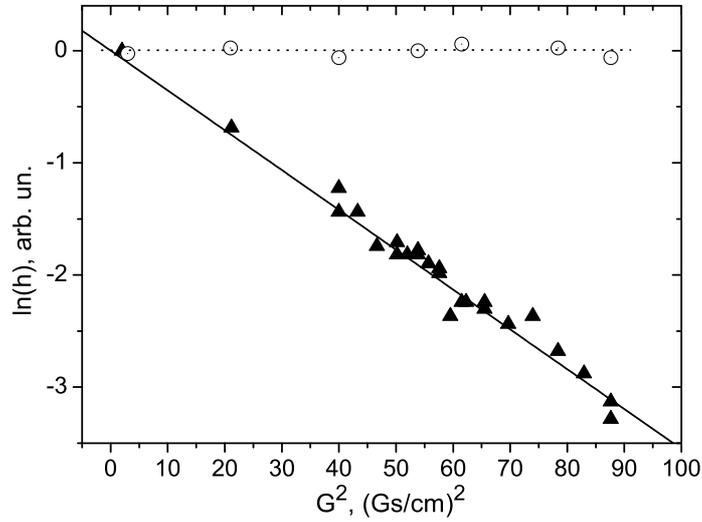}}
                \caption{\small{Diffusion decay of spin-echo in good annealed and slowly cooled hcp crystal, 1.3~K, 35.0~bar. Empty circles -- CP data for $\tau=20$~ms; solid triangles -- $SE$ data for $\tau_1=70$~ms, $\tau_2=20$~s, $D_h=8.7 \cdot 10^{-10}$~cm$^2$/s; the dotted line -- fitting to Eq.~(\ref{1}); the solid line -- fitting to Eq.~(\ref{2}).}}
                 \label{fig_good}
\end{figure}

After that this crystal was heated up to melting curve and, after annealing procedure described above (Sec.~\ref{Method}), to avoid a thermal shock, was slowly cooled down to 1.3~K (the hcp region). The results are presented in Fig.~\ref{fig_good}.
Both the absence of visible $h(G^2)$ functional dependence (see Eq.~(\ref{1})) which should be characteristic feature for $D_l= 1 \cdot 10^{-4}$~cm$^2$/s under $\tau=20$~ms at $0<G<100$~Gs/cm$^2$ and the position (0; 0) of  the intersection point of $CP$ and $SE$ data curves are the evidences of the liquid-like diffusion absence in the crystal. 

It also should be noted that monotonous pressure decrease was observed in low-quality samples with fast diffusion coefficient. The typical pressure relaxation times were about $0.5-2$~hour. After annealing of such samples along with fast diffusion process disappearing, monotonous pressure decreasing was also stopped. This relaxation indirectly confirms our speculation about liquid-like inclusions quenched from the melting curve in the samples without any annealing.
Detailed study of pressure relaxation in quenched samples is projected. 

\section{Conclusions}
It is shown that under rapidly cooling from the melting curve (without annealing) solid helium samples contain liquid-like inclusions identified by additional fast diffusion decay of echo-signal. Subsequent annealing of these samples leads to fast diffusion disappearing which is connected with crystallization of liquid-like inclusions. Coming out of these defects is accompanied by pressure relaxation in the system.

\section*{Acknowledgements}
We thank B.Cowan for useful consultations and for applying of his NMR spectrometer. This work has also been partially supported by Grant STCU \#3718, Program of Cooperation in Research and Education in Science and Technology for the 2008 Ukrainian Junior Scientist Research Collaboration, and the Ministry of Education and Science of Ukraine (Project M/386-2009).



\begin{thebibliography}{99}

\bibitem{Balibar.2008} S. Balibar, F. Caupin, {\it J. Phys: Cond. Mat.} \textbf{20}, 173201 (2008).

\bibitem{Prokofev.2007} N. Prokof'ev, {\it Adv. Phys.}  \textbf{56}, 381 (2007).

\bibitem{Boninsegni.2006} M. Boninsegni, N. Prokof'ev, and B. Svistunov, {\it Phys. Rev. Lett}  \textbf{96}, 105301 (2006).

\bibitem{Sasaki.2006} S. Sasaki, R. Ishigoro, F. Caupin, H.J. Maris, and S. Balibar, {\it Science} \textbf{313}, 1098 (2006).

\bibitem{Balatsky.2007} A.V. Balatsky, M.J. Graf, Z. Nussinov, and S.A. Trugman,  {\it Phys. Rev. B} \textbf{75}, 094201 (2007).

\bibitem{Grigorev.2007} V.N. Grigorev, V.A. Maidanov, V.Yu. Rubanskii, S.P. Rubets, E.Ya Rudavskii, A.S. Rybalko, Ye.V. Syrnikov, and V.A.Tikhii,  {\it Phys. Rev. B} \textbf{76}, 224524 (2007).

\bibitem{Balibar2.2008} S. Balibar and F. Caupin,  {\it Phys. Rev. Lett} \textbf{101}, 189601 (2008).

\bibitem{Ray.2008} M.W. Ray and R.B. Hallock, {\it Phys. Rev. Lett}  \textbf{100}, 235301 (2008).

\bibitem{Birchenko.2006} A. Birchenko, Ye. Vekhov, N. Mikhin, A. Polev, E. Rudavskii, {\it Low Temp. Phys.} \textbf{32}, 1118 (2006).

\bibitem{Carr.1954} H.Y. Carr and E.M. Purcell, {\it Phys. Rev.} \textbf{94}, 630 (1954).

\bibitem{Hann.1950} E.L. Hann, {\it Phys. Rev.} \textbf{80}, 580 (1950).

\bibitem{Garwin.1959} R.L. Garwin and H.A. Reich, {\it Phys. Rev.}  \textbf{115}, 1478 (1959).

\bibitem{Allen.1982} A.R. Allen, M.G. Richards, and J. Schratter, {\it J. Low Temp. Phys.} \textbf{47}, 289 (1982).

\bibitem{Mikheev.1983} V.A. Mikheev, N.P. Mikhin, V.A. Maidanov, {\it Sov. J. Low Temp. Phys.} \textbf{9}, 465 (1983).

\bibitem{Caupin.2008} S. Sasaki, F. Caupin, and S. Balibar, {\it Proceeding of LT-25, Amsterdam}, 275  (2008).

\bibitem{Sasaki.2008} S. Sasaki, F. Caupin, and S. Balibar, {\it Proceeding of LT-25, Amsterdam}, 146 (2008).

\bibitem{Mikhin.2001} N. Mikhin, A. Polev, and E. Rudavski, {\it JETP Letters} \textbf{73}, 470 (2001).

\bibitem{Polev.2002} A. Polev, N. Mikhin, and E. Rudavskii, {\it J. Low Temp. Phys.} \textbf{127}, 279 (2002).






\end{thebibliography}
\end{document}